\documentclass[a4paper,11pt]{article}

\usepackage[a4paper,left=3cm,right=3cm,top=3cm,bottom=2.5cm]{geometry}
\usepackage{graphicx} 
\usepackage{authblk}
\usepackage{color} 
\usepackage{amsmath}
\usepackage{amssymb} 
\usepackage{hyperref} 
\hypersetup{
	colorlinks=true,
	citecolor=black,
	filecolor=black,
	linkcolor=blue,
	urlcolor=blue
}
\usepackage[labelfont=bf]{caption}
\usepackage{dsfont}

\newcommand{\I}{\text{i}}

\title{New non-supersymmetric flux vacua in string theory}
\author[a,b]{S. Krippendorf}
\author[c]{A. Schachner}

\affil[a]{\footnotesize Arnold Sommerfeld Center for Theoretical Physics, Ludwig-Maximilians Universität, Theresienstr.~37, 80333 München, Germany}
\affil[b]{\footnotesize Universitäts-Sternwarte, Fakultät für Physik, Ludwig-Maximilians Universität,
Scheinerstr.~1, 81679 München, Germany}
\affil[c]{\footnotesize Department of Physics, Cornell University, Ithaca, NY 14853, USA}
\date{August 2023}

\begin{document}
\begin{flushright}
LMU-ASC 30/23
\end{flushright}
{\let\newpage\relax\maketitle}

\begin{abstract}
    In this note we construct large ensembles of supersymmetry breaking solutions arising in the context of flux compactifications of type IIB string theory. This class of solutions was previously proposed in~\cite{Saltman:2004sn} for which we provide the first explicit examples in Calabi-Yau orientifold compactifications with discrete fluxes below their respective tadpole constraint. As a proof of concept, we study the degree 18 hypersurface in weighted projective space $\mathbb{CP}_{1,1,1,6,9}$. Furthermore, we look at 10 additional orientifolds with $h^{1,2}=2,3$. We find several flux vacua with hierarchical suppression of the vacuum energy with respect to the gravitino mass. These solutions provide a crucial stepping stone for the construction of explicit de Sitter vacua in string theory. Lastly, we also report the difference in the distribution of $W_0$ between supersymmetric and non-supersymmetric minima.
\end{abstract}

\section{Introduction}

Breaking supersymmetry in a controlled way in string theory is a crucial step towards understanding its connection to our observable Universe. In this note, we identify new supersymmetry breaking solutions within the complex structure sector of type IIB string theory. Their contribution to the vacuum energy is positive, thereby potentially realising the uplifting mechanism proposed in~\cite{Saltman:2004sn}. At a very basic level, these solutions do not require the presence of additional sources like anti D3-branes to break supersymmetry.

For our analysis, we utilise recent numerical advances reported in~\cite{Dubey:2023dvu} to efficiently search for such minima in the large {\it string dataset} of Calabi-Yau compactifications. To access this regime, we extend the modular numerical approach of~\cite{Dubey:2023dvu} to tackle a slightly modified optimisation procedure by searching directly for extrema of the scalar potential rather than for vanishing covariant derivatives $D_I W=0$. As anticipated, this novel method is able to identify both supersymmetric\footnote{By abuse of terminology, we call solutions to $D_I W=0$ supersymmetric even if $\langle W\rangle\neq0$ at the minimum and hence, the solutions do not lift to superymmetric solutions of the full theory.} and non-supersymmetric minima.

Such minima are important not only to understand the variety of supersymmetry breaking scenarios, but also to understand mechanisms to `uplift' AdS solutions to dS. A particular variant of the original proposal~\cite{Saltman:2004sn} is `winding uplift' \cite{Hebecker:2020ejb}, see also \cite{Carta:2021sms}, which achieves exponentially small uplifts through flux choices with perturbatively flat directions.

As a proof of concept, we consider the large complex structure (LCS) regime of Type~IIB moduli stabilisation. We first report minima for the well-studied example of degree 18 hypersurface in weighted projective space $\mathbb{CP}_{1,1,1,6,9}$. Previously non-supersymmetric minima for this example were only obtained in the continuous flux approximation~\cite{Gallego:2017dvd} and the discussion on its extension to the discrete flux case seems rather vague.
Here we fill this gap by explicitly providing large ensembles of solutions in the string landscape with discrete fluxes. We will compare the resulting $W_0$ distributions for these new vacua with their supersymmetric counterparts previously discussed in \cite{Ebelt:2023clh}.

Subsequently, we study an additional $10$ geometries with two and three moduli for which we collect 250,235 solutions. In these models, we are able to demonstrate that hierarchical suppression of the supersymmetry breaking scale is easily achieved in large samples of vacua. Our ensemble provides a first glance at the broad landscape of non-supersymmetric solutions and a golden opportunity to compare their statistical properties against those for supersymmetry preserving vacua.

The rest of this note is organised as follows. In Section~\ref{sec:theory} we describe the type of solution we are searching for. In Section~\ref{sec:numerics} we provide our numerical results and we conclude in Section~\ref{sec:conclusions}.

\section{Class of supersymmetry breaking solutions}
\label{sec:theory}

We are interested in the effective field theory for the complex structure sector of Type~IIB string theory in Calabi-Yau orientifold compactifications. In this setting, the action for the complex structure moduli $Z^i$ and the axio-dilaton $\tau$ is determined by the K\"ahler and flux superpotential
\begin{eqnarray}
K(Z^i,\overline{Z}^i,\tau,\overline{\tau})&=&-\log{\left(-\I\, \Pi^\dagger(\overline{Z}^i)\cdot \Sigma\cdot \Pi(Z^i)\right)}-\log{\left(-\I(\tau-\bar{\tau})\right)}\,\label{eq:kaehler} ,~\\[0.2em]
W(Z^i,\tau)&=&(f-\tau h)\cdot\Sigma\cdot \Pi(Z^i)~. \label{eq:superpotential}
\end{eqnarray}
Here, $f,h\in \mathbb{Z}^{2(h^{1,2}+1)}$ are integer flux vectors arising from integrating the Type~IIB NSNS and RR 3-forms $H_3,F_3$ over 3-cycles. The period vector $\Pi$ is determined by the prepotential $F$ through
\begin{equation}
    \Pi(Z^i)=\left(
            \begin{matrix}
                2F-Z^i F_i\\
                F_i\\
                1\\
                Z^i
            \end{matrix}
        \right)\; ,\quad \Sigma = \left (\begin{array}{cc}
0 & \mathds{1} \\ [-0.2em]
-\mathds{1} & 0
\end{array} \right )\, .
\end{equation}
In this note, we work in the LCS regime where the pre-potential can be written as \cite{Morrison:1991cd,Hosono:1994av,Hosono:1994ax}
\begin{eqnarray}
F=-\dfrac{1}{6}\kappa_{ijk} \,  Z^i\,  Z^j \,  Z^k +  \frac{1}{2} \,{a_{ij} \,  Z^i\,  Z^j} +  \,{b_{i} \,   Z^i} +  \frac{\I}{2} \, \,{\tilde \xi} -\dfrac{1}{(2\pi \I)^{3}}\sum_{q}\, n_{q}^{(0)}\, \text{Li}_{3}\left (\mathrm{e}^{2\pi \I\, q_i\, Z^i}\right )\, .
\end{eqnarray}
Here, we follow the conventions of~\cite{Dubey:2023dvu} to which we refer for more details. We simply note that all of the relevant input parameters can be readily computed using mirror symmetry, see \cite{Demirtas:2023als} for a recent discussion.

This work is concerned with finding local minima of the standard no-scale supergravity scalar potential for $Z^i,\tau$ induced by fluxes given by
\begin{equation}\label{eq:scalar_potential}
    V=e^K\, K^{I\bar{J}}\, D_I W  \overline{D_{J}W}\, , \quad D_I W=\partial_I W+W\partial_I K ~\, .
\end{equation}
Here, $K^{I\bar{J}}$ is the inverse K\"ahler metric and $D_I W$ denotes the covariant derivative of the superpotential. To find minima, we need to identify critical points $\partial_I V=0$ and subsequently perform additional checks such as positive definiteness of the Hessian.

In a previous work \cite{Dubey:2023dvu}, we considered solutions satisfying the $F$-flatness conditions $D_I W=0$ which, looking at \eqref{eq:scalar_potential}, lead to Minkowski vacua because $\langle V\rangle \equiv 0$. In this note, we instead focus on solutions to $\partial_I V=0$ satisfying $D_I W\neq 0$. Our main motivation is rooted in \cite{Saltman:2004sn} arguing that this class of solutions can potentially be useful for dS model building given that the vacuum energy $V_0=\langle V\rangle>0$ is strictly positive. For a fully fledged realisation of the mechanism of \cite{Saltman:2004sn}, one would also need to take into account Kähler moduli which will be studied in future works. Here, we simply comment on the fact that, in order to uplift AdS minima in either KKLT~\cite{Kachru:2003aw} or LVS~\cite{Balasubramanian:2005zx} generically demands slightly different uplift energies which result in additional constraints on these supersymmetry breaking solutions. Indeed, cancelling the respective AdS scales for KKLT and LVS to first approximation, one requires slightly different values of $V_0$ for a successful uplift to de Sitter with small cosmological constant
\begin{equation}
    V_0\simeq|W_0|^2\text{ for KKLT}\; ,\quad V_0\simeq\dfrac{|W_0|^2}{\mathcal{V}}\text{ for LVS}\, .
\end{equation}
Below, we find that the milder hierarchies for LVS can already be found in our ensemble of solutions.

\section{Numerical results}
\label{sec:numerics}

To demonstrate our modified search algorithm for non-SUSY vacua, we show numerical results for $11$ models with two- and three-complex structure moduli taking into account instanton contributions up to degree 10. We include the well-known example of $\mathbb{CP}_{1,1,1,6,9}[18]$ for which the prepotential has been commonly discussed in the literature. Alongside this paper, we provide the relevant details for the other geometries in the ancillary files and sample solutions for $\mathbb{CP}_{1,1,1,6,9}[18]$ in Appendix~\ref{app:p11169}.

Besides the difference in the optimisation module, we also have to carefully select our sampling method for respective starting points of the moduli and the flux values. In particular, the ${\rm ISD}_{\pm}$ sampling strategies introduced in \cite{Dubey:2023dvu} lead to initialisations close to supersymmetric minima to which our algorithm converges quickly rather than exploring parameter space regions where non-supersymmetric minima are located. Hence, we use our random initialisation for starting points and flux vectors. As pointed out before, this choice reduces the success rate of finding such vacua. Nevertheless, we are able to identify reasonably large ensembles of solutions with supersymmetry breaking.

\subsection{Vacua for $\mathbb{CP}_{1,1,1,6,9}[18]$}

\begin{figure}[t!]
    \centering
    \includegraphics[width=0.75\textwidth]{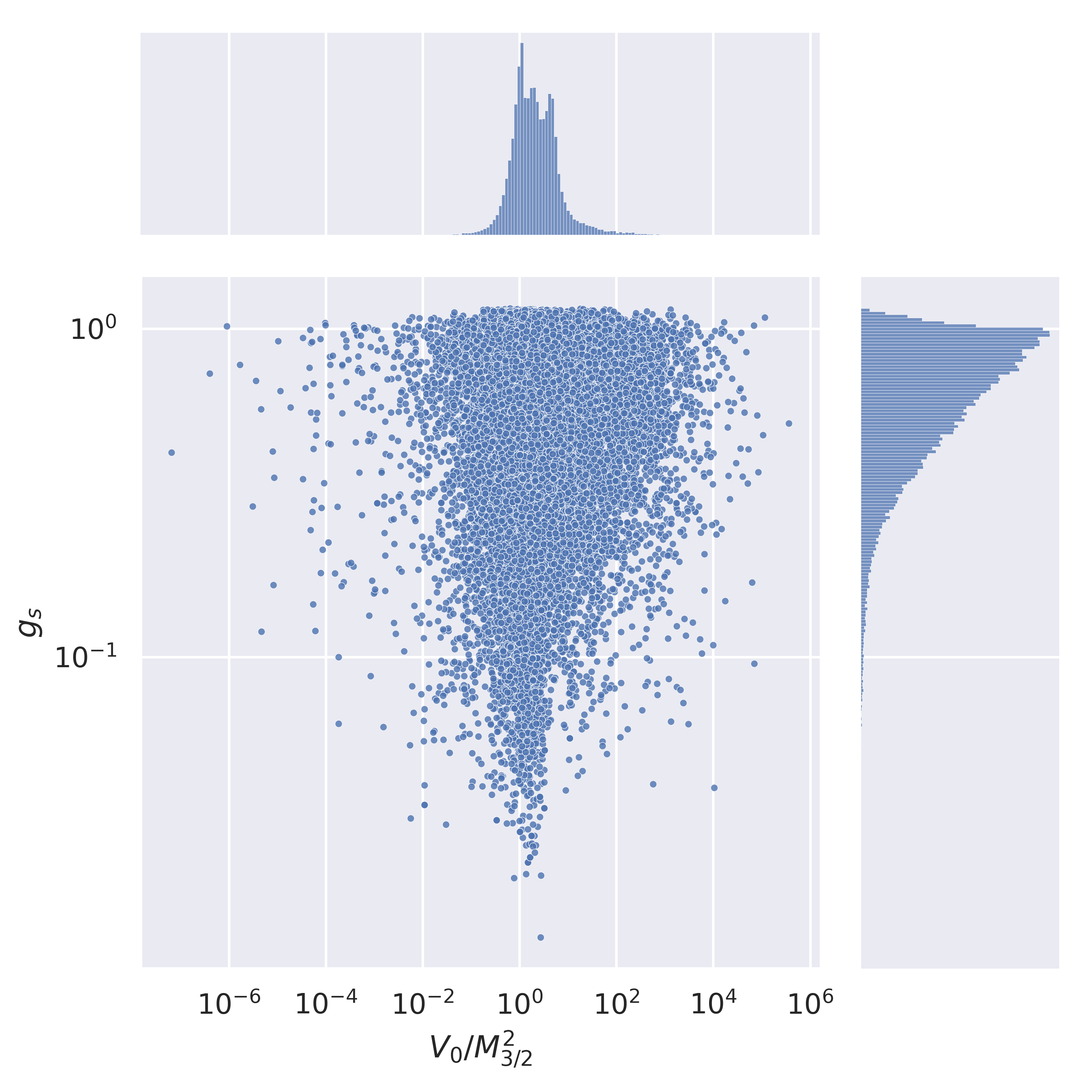}
    \caption{We show the distribution of vacuum energies in units of gravitino mass and the string coupling $g_s$ for $\mathbb{CP}_{1,1,1,6,9}[18]$.}
    \label{fig:distribution}
\end{figure}

\begin{figure}[t!]
    \centering
    \includegraphics[width=1.\textwidth]{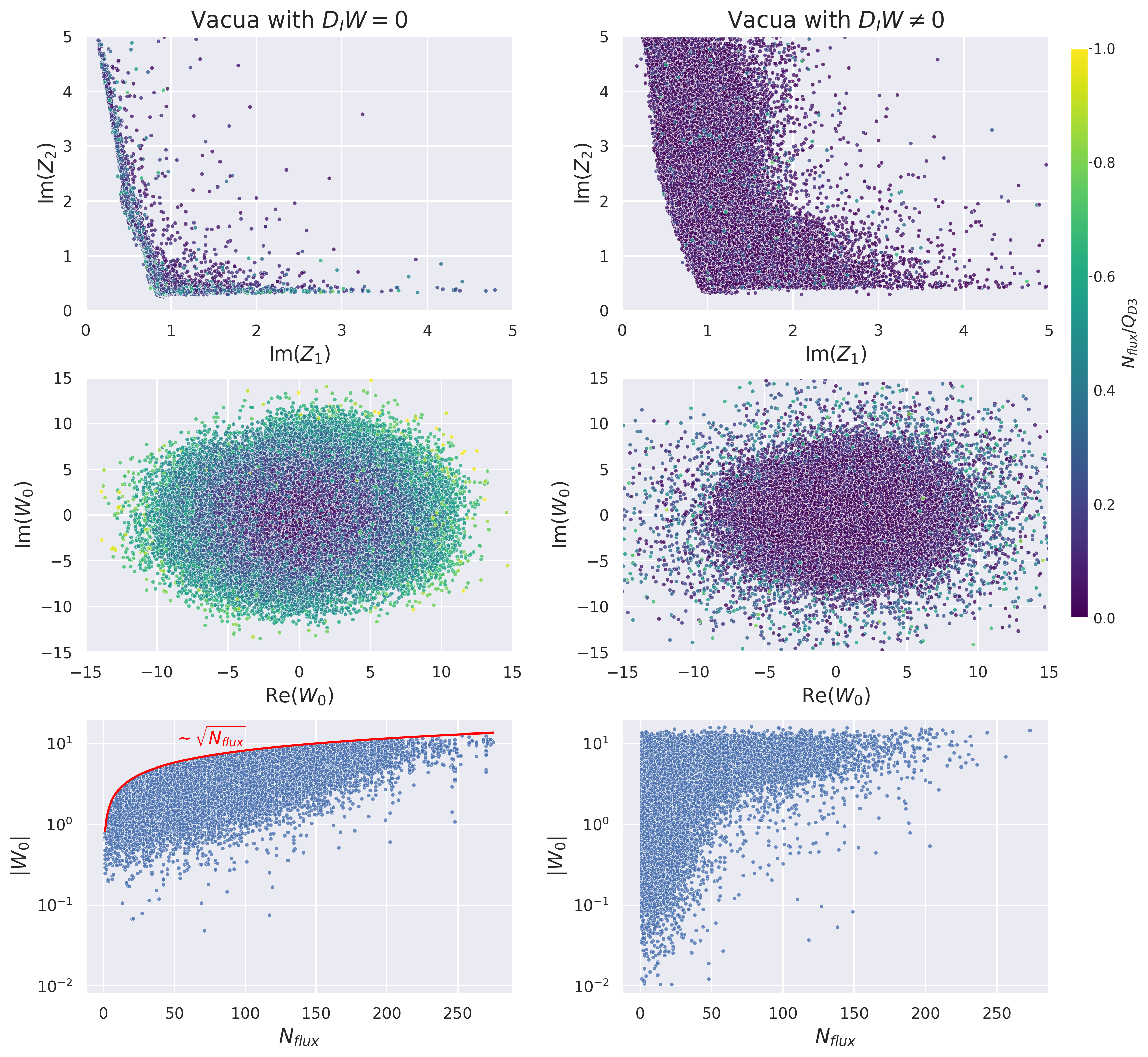}
    \caption{Confronting supersymmetric and non-supersymmetric solutions for $\mathbb{CP}_{1,1,1,6,9}[18]$.
    \emph{Top:} Comparison of the imaginary parts of moduli expectation values between SUSY and non-SUSY vacua.
    \emph{Middle:} Comparing the distributions of $W_0$ for SUSY and non-SUSY vacua.
    \emph{Bottom:} $|W_0|$ as a function of $N_{\text{flux}}$ for both types of minima.
    }
    \label{fig:w0distributions}
\end{figure}

First, let us look at the example of $\mathbb{CP}_{1,1,1,6,9}[18]$ for which the relevant orientifold data was summarised e.g.~in \cite{Louis:2012nb}. Using the algorithm outlined above, we collected 77,849 vacua with $D_IW\neq 0$. To allow for further analytic inspection of our solutions, we list selected candidates with small $V_0$ in Appendix~\ref{app:p11169}. 

The distribution of our solutions and their vacuum energy $V_0$ in units of the gravitino mass is shown in Figure~\ref{fig:distribution} where the latter is calculated using the Kähler and superpotential from Eqs.~\eqref{eq:kaehler} and~\eqref{eq:superpotential}, i.e.,~$M_{3/2}=e^{K/2}|W|$. We note that in our ensemble we identify examples where the ratio of $V_0/M_{3/2}^2$ is hierarchically suppressed. These solutions are generally speaking interesting because such hierarchical suppressions allow for room for additional control in the EFT and are generically required for a successful de Sitter uplift (cf.~Sec.~\ref{sec:theory}). Such additional protection is strictly speaking not necessary for the sector which we focus on in this work, i.e.,~the EFT derived by GPK~\cite{Giddings:2001yu} is also valid off-shell and it is safe to use this theory also near non-SUSY minima. Having said this, when including K\"ahler moduli and perturbative corrections, the validity of this EFT has to be justified more carefully.

Next, let us compare our new solutions against their more established counterparts satisfying $D_I W=0$. To this end, we collected an additional 77,849 of solutions to $D_I W=0$ by using the same sampling algorithm, namely uniformly sampling fluxes and initial guesses (see also~\cite{Martinez-Pedrera:2012teo}). Looking at the distribution of moduli vacuum expectation values in the top row of Fig.~\ref{fig:w0distributions}, we find that the SUSY breaking minima are located further inside the K\"ahler cone in comparison to the SUSY solutions when using the random sampling method. While we typically expect these distributions to depend on the chosen sampling methods,\footnote{See~\cite{Dubey:2023dvu} for a comparison of sampling strategies in the context of SUSY solutions.} we stress again that the sampling of input fluxes and initial guesses is exactly the same for both types of solutions. Hence, the resulting distributions show structures mainly inherited from the special class of solutions. In particular, in the case of $D_I W=0$ solutions, the ISD condition leaves an imprint on the moduli VEVs by localising the latter close to the boundary of moduli space. This needs to be contrasted with $D_I W\neq 0$ which are broadly scattered across the Kähler cone.

In fact, we can make similar observations by comparing the distribution of the VEV of the superpotential $W_0=\sqrt{2/\pi}\langle e^{K/2}W\rangle $ between SUSY and non-SUSY minima as shown in the middle row of Fig.~\ref{fig:w0distributions}. As pointed out in \cite{Ebelt:2023clh}, the $D_I W=0$ solutions are to first approximation Gaussian distributed with the standard deviation proportional to $\sqrt{N_{\text{flux}}}$. Here, this can be seen more clearly in the bottom row of Fig.~\ref{fig:w0distributions} which shows the distribution of $|W_0|$ as a function of $N_{\text{flux}}$. While the plot on the right shows almost no persistent correlations, the one on the left is perfectly bound by $\approx 0.82\cdot\sqrt{N_{\text{flux}}}$. While the analysis in \cite{Ebelt:2023clh} used ISD biased sampling as introduced in \cite{Dubey:2023dvu}, we sample here all of the input parameters (fluxes and initial guesses) from uniform distributions. Hence, this reinforces our claim that the observed Gaussian behaviour with $\sigma \sim \sqrt{N_{\text{flux}}}$ of the $W_0$ distribution is largely sampling independent. In contrast, the corresponding $W_0$ distribution for $D_I W\neq 0$ vacua is clearly much less constrained. Indeed, the ISD conditions impose stringent constraints on the $F$-flat vacua\footnote{Thus, some of the structures that have been observed in the distributions of vacua like in \cite{Dubey:2023dvu,Ebelt:2023clh} should not necessarily be attributed to string theory compactifications alone, but rather to restricting to SUSY solutions.}, while general solutions with $D_I W\neq 0$ can come in a variety of different incarnations. These differences can in the future be analysed more thoroughly using our methods.

\subsection{Analysis for 10 models at $h^{1,2}=2,3$}

\begin{table}[t!]
    \centering
    \begin{tabular}{c|c|c|c}
         $h^{1,1}$ & $h^{1,2}$ & $Q_{D3}$ & $\sharp$vacua  \\
         \hline
         \hline
          144& 2 & 148  & 17,369 \\
         \hline
          120 & 2 & 124  &55,068 \\
         \hline
          132 & 2 & 136  & 32,340 \\
         \hline
          128 & 2 & 132 & 32,771\\
         \hline
          272 & 2 & 276 & 33,619\\
    \end{tabular}
    \hspace{0.5cm}
    \begin{tabular}{c|c|c|c}
         $h^{1,1}$ & $h^{1,2}$ & $Q_{D3}$ & $\sharp$vacua  \\
         \hline
         \hline
          99 & 3 & 104  & 656 \\
         \hline
          115 & 3 & 120  & 10,859 \\
         \hline
          107 & 3 & 112 & 53,158 \\
         \hline
          119 & 3 & 124 & 238\\
         \hline
          243 & 3 & 248 & 14,157\\
    \end{tabular}
    \caption{Summary of our compactification manifolds with their respective Hodge numbers, tadpole values and number of vacua. In total, we find $250,235$ solutions.}
    \label{tab:models}
\end{table}

\begin{figure}[t!]
    \centering
    \includegraphics[width=0.75\textwidth]{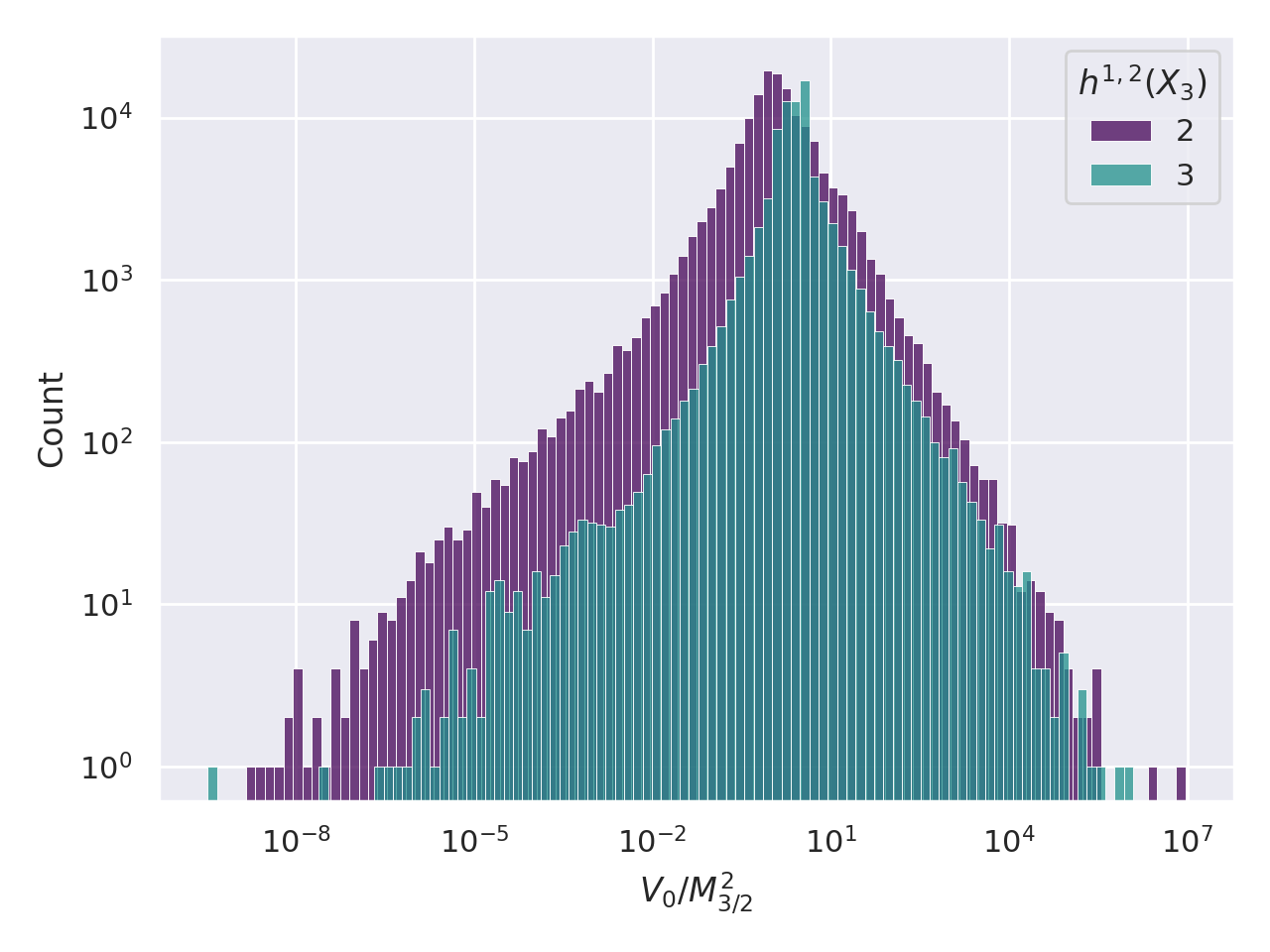}
    \caption{We show the scatterplot of values of $|W_0|$ and vacuum energies in units of gravitino mass for the geometries in Tab.~\ref{tab:models}.}
    \label{fig:distribution23}
\end{figure}

\begin{figure}[t!]
    \centering
    \includegraphics[width=1.\textwidth]{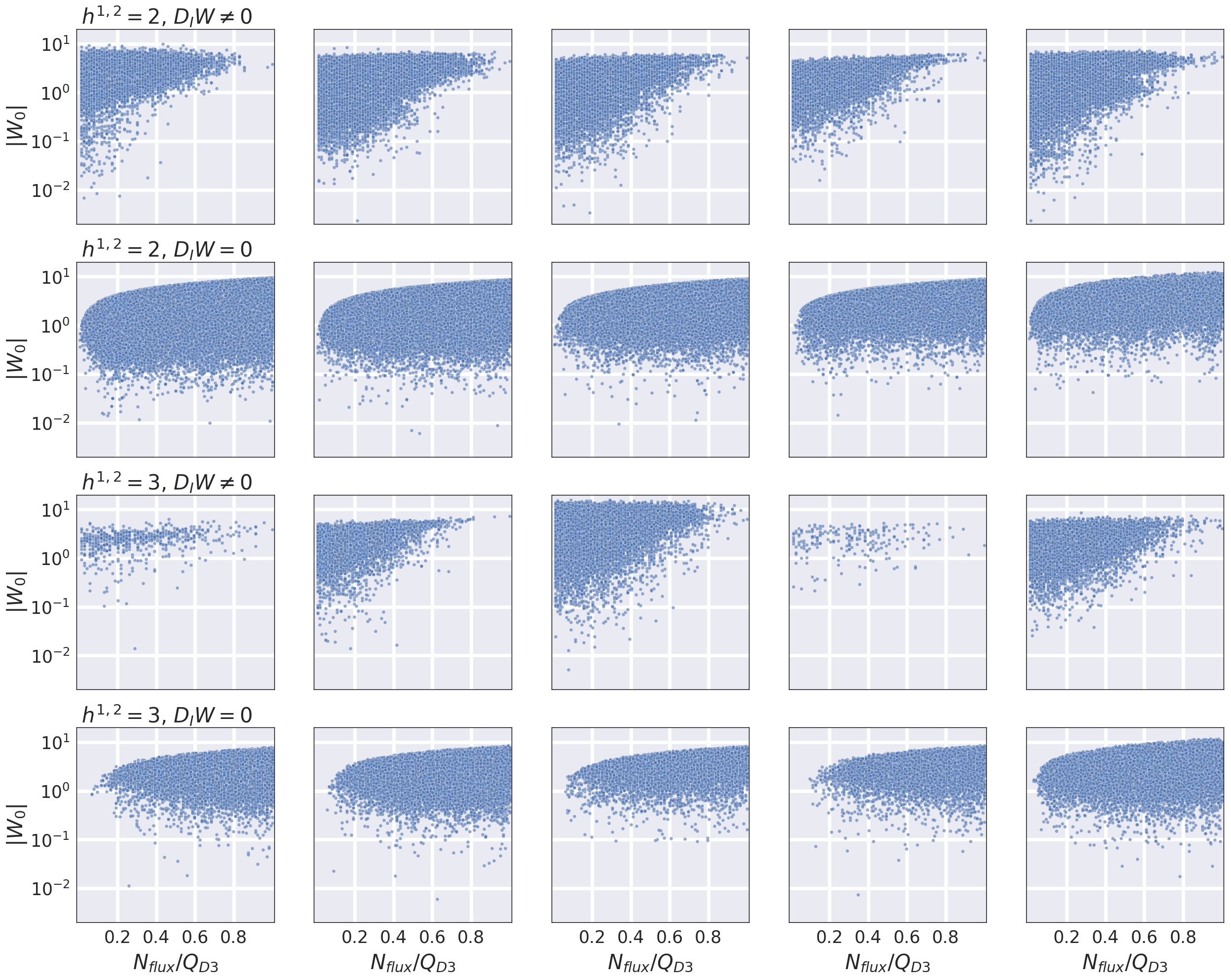}
    \caption{Different structures in the distributions of minima: We plot the absolute value of $W_0$ as a function of $N_{\text{flux}}/Q_{D3}$ for our solutions with $D_I W\neq 0$ in the first and third row. For a direct comparison, we show the same plots for solutions to $D_I W= 0$ previously constructed in \cite{Ebelt:2023clh} in the second and fourth row.}
    \label{fig:distributionW023NS}
\end{figure}

Lastly, let us study a larger sample of solutions across various different orientifold models. To this end, we selected in total 10 smooth CY orientifolds from the database of \cite{Crino:2022zjk} with two and three complex structure moduli. We summarise our results in Tab.~\ref{tab:models} where find in total 250,235 flux vacua with $D_I W\neq 0$. Generally speaking we note that finding these non-supersymmetric minima currently appears computationally more expensive, but clearly not prohibitive. The distribution of the vacuum energy $V_0$ in units of the squared gravitino mass $M_{3/2}^2$ is shown in Fig.~\ref{fig:distribution23}. Similar to Fig.~\ref{fig:distribution}, we observe several vacua with hierarchically suppressed vacuum energy which could potentially play a role in de Sitter uplifting.

Lastly, we show the behaviour of the absolute value of $W_0$ as a function of $N_{\text{flux}}/Q_{D3}$ in Fig.~\ref{fig:distributionW023NS}. To allow for a direct comparison, we show the same plots for 1,466,383 solutions with $D_IW=0$ previously constructed in \cite{Ebelt:2023clh}. As observed before in Fig.~\ref{fig:w0distributions}, $D_I W=0$ solutions show the same characteristic behaviour $\sigma\sim \sqrt{N_{\text{flux}}}$ across all models. In contrast, SUSY-breaking vacua with $D_I W\neq 0$ exhibit no noteworthy dependence on $N_{\text{flux}}$ in our ensemble. These differences will be analysed in more detail in future works.

\section{Conclusions}
\label{sec:conclusions}

To characterise the low energy physics coming from string theory, it is imminent to study its solutions as broadly as possible. In this quest, we find numerical techniques crucial as they allow us to construct such vacua using efficient implementations of a combination of standard algorithms. In this note, we demonstrated that this approach can be extended to the search for supersymmetry breaking solutions. In particular, solutions of this type were proposed long ago, but so far only available in toroidal examples or in the continuous flux approximation. Here, we were able to work with explicit Calabi-Yau orientifolds and to obtain large ensembles of such supersymmetry breaking solutions using only moderate computational resources.

We find several solutions where the scale of the vacuum energy is suppressed compared to the gravitino mass, a feature which is desirable in the construction of hierarchically small uplifts. For instance such suppressed uplifts would be needed in the LARGE Volume Scenario~\cite{Balasubramanian:2005zx}.

Clearly, this work only served as a proof of concept for future, more detailed investigations of supersymmetry breaking vacua. Most prominently, in the quest for de Sitter minima, it is important to include Kähler moduli stabilisation. With our modular approach, we will be able to include this aspect in the future. Another line of further investigation is to acquire more empirical solutions to compare the observed statistical distribution with existing approximations to these distributions in the literature~\cite{Denef:2004cf}. Finally, it would be interesting to compare the scales in our supersymmetry breaking solutions with those appearing in other approaches such as in~\cite{DeLuca:2021pej}.

\subsection*{Acknowledgments}

We would like to thank Michele Cicoli and Pellegrino Piantadosi for discussions and collaborations on related topics. We are furthermore grateful to Liam McAllister and Alexander Westphal for fruitful discussions.
We especially thank Andres Rios-Tascon for providing the code to compute GV and GW invariants. AS thanks the Ludwig Maximilian University of Munich and ICISE in Quy Nhon, Vietnam, for hospitality where parts of this work have been completed. The research of AS is supported by NSF grant PHY-2014071.

\appendix

\section{Example solutions for $\mathbb{CP}_{1,1,1,6,9}[18]$}
\label{app:p11169}

\begin{table}[h!]
\centering
\begin{tabular}{|c||c|c|c|c|c|c|c|c|c|c|c|c||c|}
\hline 
 &  &  &  &  &  &  &  &  &  &  &  & &  \\[-1.1em] 
Sol: & $f_{1}$ & $f_{2}$ & $f_{3}$ & $f_{4}$ & $f_{5}$ & $f_{6}$ & $h_{1}$ & $h_{2}$ & $h_{3}$ & $h_{4}$ & $h_{5}$ & $h_{6}$& $N_{\text{flux}}$ \\ [0.2em]
\hline 
\hline 
 &  &  &  &  &  &  &  &  &  &  &  &  & \\[-1.1em] 
a & 0& 18&  0&  1&  1&  3& 0 & $-7$& $-1$& 0&  1& $-2$ &28 \\ [0.2em]
\hline 
 &  &  &  &  &  &  &  &  & &  &  &  &  \\[-1.1em] 
b & 0 & 1 & 1 & 1 & $-3$ & 7 & 0 & $-3$ & $-5$ & 0 & 1 & $-3$ &24 \\[0.2em]
\hline 
 &  &  &  &  &  &  &  &  &  &  &  &&   \\[-1.1em] 
c & 0 & 18 & 3 & 2 & 4 & $-4$ & 0 & $-7$ & 0 & 0 & 0 & 1 & 31\\[0.2em]
\hline 
 &  &  &  &  &  &  &  &  &  &  &  & &  \\[-1.1em] 
d & 0 & 8 & $-3$ & 1 & 1 & 3 & 0 & $-4$ & $-2$ & 0 & 2 & $-5$ & 41\\[0.2em]
\hline 
 &  &  &  &  &  &  &  &  &  &  &  &  & \\[-1.1em] 
e & 0 & 11 & 5 & 0 & 2 & $-4$ & 0 & 4 & 4 & $-1$ & 2 & $-2$ & 20\\[0.2em]
\hline 
 &  &  &  &  &  &  &  &  &  &  &  & &  \\[-1.1em] 
f & 0 & 6 & $-1$ & $-1 $& $-6$ & 8 & 0 & 16 & $-2$ & 2 & 1 & 4& 114  \\[0.2em]
\hline 
 &  &  &  &  &  &  &  &  &  &  &  & &  \\[-1.1em] 
g & 0 & $-10$ & $-2$ & $-1$ & 0 & 1 & 0 & $-4$ & $-3$ & 1 &$ -2$ & 2 & 19\\[0.2em]
\hline 
\end{tabular} 
\caption{Flux vectors of some of our solutions of $\mathbb{CP}_{1,1,1,6,9}[18]$. The individual solutions can be found in Tab.~\ref{tab:solutions}}\label{tab:fluxes}
\end{table}
\begin{table}[h!]
\centering
\resizebox{\columnwidth}{!}{\begin{tabular}{|c||c|c|c|c|c|c|}
\hline 
 &  &  &  &  &  &         \\[-1.1em] 
Sol: &  $Z_{1}$ & $Z_{2}$ &  $\tau$& $g_{s}$ & $W_{0}$ &    $V_{0}$ \\ [0.2em]
\hline 
\hline 
 &  &  &  &  &  &         \\[-1.1em] 
a& $0.71+1.04\I $ & $-0.69+2.26\I$  & $-0.25+1.80\I $ & 0.56 & $-46.53 + 42.09\I$ &  $7.76\cdot 10^{-4}$ \\[0.2em]
\hline 
 &  &  &  &  &  &         \\[-1.1em] 
b& $-3.17+2.54\I$  &$ 5.31+1.06\I $& $0.19+3.35\I $ & 0.30 & $-56.89 +72.18\I $&  $9.63\cdot 10^{-3}$ \\[0.2em]
\hline 
 &  &  &  &  &  &     \\[-1.1em] 
c& $1.19+1.01\I $ & $2.94+3.86\I $ & $-0.40+5.30\I$  & 0.19 & $27.11 + 119.94\I $&  $4.24\cdot 10^{-3}$ \\[0.2em]
\hline 
 &  &  &  &  &  &         \\[-1.1em] 
d& $2.04+1.79\I $ & $1.92+1.38\I $ & $-0.35+2.24\I$  & 0.45 & $26.01 + 76.43 \I$ &  $1.16\cdot 10^{-3}$ \\[0.2em]
\hline 
 &  &  &  &  &  &         \\[-1.1em] 
e& $-1.71+1.21\I$  & $-0.02+1.16\I$  & $-0.03+1.07\I$  & 0.94 & $-31.49  -9.98 \I$ &  $3.25\cdot 10^{-4}$ \\[0.2em]
\hline 
 &  &  &  &  &  &          \\[-1.1em] 
f& $1.81+1.06\I $ & $-1.76+1.26\I $ & $0.09+1.60\I $ & 0.62 & $81.75 -20.23 \I $&  $6.66\cdot 10^{-3}$\\[0.2em]
\hline 
 &  &  &  &  &  &        \\[-1.1em] 
g& $-1.00+1.10\I $ & $0.93+1.29\I $ & $-0.40+1.14\I $ & 0.88 & $28.13 -24.31 \I$ &  $3.39\cdot 10^{-3}$ \\[0.2em]
\hline 
\end{tabular} }
\caption{The properties of our solutions for the flux vectors (cf.~Tab.~\ref{tab:fluxes}) for  $\mathbb{CP}_{1,1,1,6,9}[18]$.}\label{tab:solutions}
\end{table}

In this appendix, we provide a few example solutions for the geometry $\mathbb{CP}_{1,1,1,6,9}[18]$ where the prepotential has been studied extensively in the literature~\cite{Martinez-Pedrera:2012teo,Demirtas:2019sip}. The flux vectors and the associated solutions are listed in tables~\ref{tab:fluxes} and \ref{tab:solutions} respectively. We stress that the units for the moduli quoted here are not canonically normalised. We checked the minimum up to degree $100$ in the instanton expansion.

\newpage

\bibliographystyle{utphys}
\bibliography{bibliography}

\end{document}